\documentclass[a4paper]{article}

\usepackage{amsmath,amssymb}
\usepackage{graphicx,epsfig}

\usepackage{amsthm}
\theoremstyle{plain} 
\newtheorem{theorem}{Theorem} 
 
\theoremstyle{definition} 
\newtheorem{definition}{Definition} 
\theoremstyle{remark}

\newcommand{\grad}{\operatorname{{\mathrm grad}}}

\begin{document}

\title{Dynamical systems in cosmology}
\author{Christian G.~B\"ohmer\footnote{c.boehmer@ucl.ac.uk (corresponding author)}~${}^{1}$ and Nyein Chan\footnote{nchan@isyedu.org}~${}^{2,3}$\\
${}^1$Department of Mathematics, University College London\\
Gower Street, London, WC1E 6BT, UK\\
${}^2$Faculty of Engineering, Computing and Science\\
Swinburne University of Technology, Sarawak Campus\\
Jalan Simpang Tiga, 93350 Kuching, Sarawak, Malaysia\\
${}^3$International School Yangon\\
Shwe Taungyar Street, Yangon, Myanmar}
\date{}
\maketitle

\begin{abstract}
  Cosmology is a well established research area in physics while dynamical systems are well established in mathematics. It turns out that dynamical system techniques are very well suited to study many aspects of cosmology. The aim of this book chapter is to provide the reader with a concise introduction to both cosmology and dynamical system. The material is self-contained with references to more detailed work. It is aimed at applied mathematics and theoretical physics graduate level students who have an interest in this exciting topic.
\end{abstract}
\mbox{}\\[3ex]
These lecture notes are based on the thesis of NC and on lectures given by CGB at the London Taught Course Centre (LTCC).

\clearpage
\tableofcontents

\clearpage
\section{A brief introduction to cosmology}

\subsection{The basics}

Cosmology is the study of the universe as a whole, and its aim is to understand the origin of the universe and its evolution. The study of the cosmos is as old as humanity and has always been fascinating. Physical cosmology\footnote{We will drop the word physical soon. It is used here to emphasise the scientific aspect of cosmology opposed to the philosophical or religious studies.} is the scientific study of the universe as a whole based on the laws of physics. The dominant interaction between macroscopic objects is the gravitational force. Therefore, we must study the dynamics of the universe within the framework of Einstein's theory of General Relativity which was formulated in 1916. In simple terms, the main concept of general relativity is the following equation
\begin{align}
  \text{geometry} = \kappa \times \text{matter}
\end{align}
where $\kappa = 8\pi G/c^4$ is a coupling constant which determines the strength of the gravitational force. $G$ is Newton's gravitational constant and $c$ is the speed of light. 

The Einstein field equations are a set of 10 coupled non-linear PDEs, or in other words, very difficult equations to deal with in general~\cite{Bruhat}. However, these equations can be simplified considerably by making some suitable assumptions. In cosmology~\cite{DodelsonModCosmo} this is known as the cosmological principle. It is an axiom which states that the universe is homogeneous and isotropic when viewed over large enough scales. 

These scales are of the order of $100 - 1000\, {\rm MPc}$. To translate this into more practical units, we note that $1\, {\rm pc} \approx 3.26\, {\rm ly} \approx 10^{12} {\rm km}$. This means $100\, {\rm MPc} \approx 326\, {\rm Mly} \approx 10^{18}\, {\rm km}$, and has the simple practical implication that we cannot test the cosmological principle directly by making observations at two points in the universe separated by cosmologically significant scales. However, there are other possibilities of testing the cosmological principle. For instance, if we were to observe a very large structure in the universe which is bigger than $100\, {\rm MPc}$, say, then this would force us to revise this number upwards. It turns out that such very large structure have already been observed, see the Clowes--Campusano Large Quasar Group for one such example.

Henceforth we assume that the cosmological principle is valid for some suitable length scale. A homogeneous and isotropic 4-dimensional Lorentzian manifold is characterised by only one function which is usually denoted by $a(t)$ and one constant $k = (\pm 1, 0)$. Such models were studied independently by Friedmann, Lema\^{\i}tre, and Robertson \& Walker. The function $a(t)$ is called the scale factor and is the only dynamical degree of freedom in the cosmological Einstein field equations. The constant $k$ characterises the curvature of the so-called constant time hypersurfaces, $k=0$ corresponds to a Euclidean space, $k=+1$ to a 3-sphere and $k=-1$ to hyperbolic space. The cosmological Einstein field equations are given by
\begin{subequations}
  \label{field1}
  \begin{align}
    3\frac{\dot{a}^2}{a^2} + 3\frac{k}{a^2} - \Lambda &= \kappa\, \rho
    \label{field1a}\\
    -2\frac{\ddot{a}}{a} - \frac{\dot{a}^2}{a^2} - \frac{k}{a^2} + \Lambda &= \kappa\, p.
  \end{align}
\end{subequations}
Here $\Lambda$ is the so-called cosmological constant, $\rho$ and $p$ are the energy density and pressure of some matter components, respectively. This matter could be a perfect fluid with prescribed equation of state, or a scalar field for instance. More complicated forms of matter can also be included.  One can verify by direct calculation that these two equations imply the energy-conservation equation
\begin{align}
  \dot{\rho} + 3\frac{\dot{a}}{a}(\rho + p) = 0.
  \label{cons1}
\end{align}
In cosmology one assumes that every matter component satisfies its own conservation equation, which does not follow from the field equations but must be assumed or derived separately. Inspection of Eqs.~(\ref{field1}) shows that we have two equations but 3 functions to be found, namely $a(t)$, $\rho(t)$ and $p(t)$. This system of equations is under-determined. In order to close it, we will assume a linear equation of state between the pressure and the energy density $p = w\rho$, where the equation of state parameter $w \in (-1,1]$. 

The scale factor $a(t)$ is a measure of the size of the universe at time $t$. However, since we do not have an absolute length scale, the numerical value of $a(t)$ can be rescaled. One convention is to choose $a(t_{\rm today})=1$ and compare the universe's size with its current value. Moreover, it turns out to be useful to introduce the Hubble function $H(t) := \dot{a}/a$ which is a measure of the universe's expansion rate at time $t$. A positive value for this quantity was first observed by Edwin Hubble in 1929, thereby giving experimental evidence to an expanding universe. Today's value $H_{\rm today}$ is of the order of $70 {\rm km}/{\rm s}/{\rm Mpc}$. 

Let us now rewrite the field equations using the Hubble parameter. Firstly, we need the relation
\begin{align}
  \dot{H} = \frac{\ddot{a}}{a} - \frac{\dot{a}^2}{a^2} = \frac{\ddot{a}}{a} - H^2
\end{align}
which allows us to write~(\ref{field1}) in the following form
\begin{subequations}
  \begin{align}
    3H^2 + 3\frac{k}{a^2} - \Lambda &= \kappa\, \rho
    \label{field2a}\\
    -2\dot{H} - 3H^2 - \frac{k}{a^2} + \Lambda &= \kappa\, p.
  \end{align}
\end{subequations}
Equation~(\ref{field2a}) is of particular interest to us. By dividing the entire equation by $3H^2$ we arrive at 
\begin{align}
    1 = \frac{\kappa\, \rho}{3H^2} + \frac{\Lambda}{3H^2} - \frac{k}{a^2 H^2}
\end{align}
and observe that each of the three terms is dimensionless. 

It is common to introduce the following dimensionless density parameters
\begin{align}
  \Omega = \frac{\kappa\, \rho}{3H^2}, \qquad
  \Omega_{\Lambda} = \frac{\Lambda}{3H^2}.
  \label{density}
\end{align}
Note that $\Omega$ may contain different forms of matter, the total matter content might contain a pressure-less perfect fluid (standard matter or sometimes called dust) and radiation, in which case one would write $\Omega = \Omega_{\rm m} + \Omega_{\rm r}$. Before getting started with dynamical systems and their application to cosmology, we need to discuss some of the well known solutions in cosmology.

\subsection{Cosmological solutions}

We will now discuss the most important solutions of the field equations~(\ref{field1}). This is needed in order to understand and interpret the solutions encountered later using dynamical systems techniques. 

In order to simplify the equations, we will assume that the spatial curvature parameter vanishes, i.e.~$k=0$ and we will also neglect the cosmological term $\Lambda = 0$. Let us firstly assume that the equation of state parameter $w=0$. This corresponds to a matter dominated universe. One can immediately integrate the conservation equation~(\ref{cons1}) and find that
\begin{align}
  \rho \propto a^{-3}.
  \label{matter}
\end{align}
This result is not unexpected since we find that density is inversely proportional to volume. Using this result in the field equation~(\ref{field1a}) yields the solutions $a(t) \propto t^{2/3}$. 

Secondly, we consider $w=1/3$ which corresponds to a radiation dominated universe. In that case, the conservation equation gives
\begin{align}
  \rho \propto a^{-4}.
  \label{rad}
\end{align}
and the remaining field equations can be solved to find $a(t) \propto t^{1/2}$. 

Lastly, we consider the case where $\rho = p = 0$, however, we assume $\Lambda >0$. Then, we can integrate~(\ref{field2a}) and find
\begin{align}
  a(t) \propto \exp\left(\sqrt{\Lambda/3}\, t\right).
\end{align}
This solution is generally called the de Sitter solution and corresponds to a universe which undergoes an accelerated expansion.

\subsection{A very brief history of the universe}
\label{sec:briefhist}

Based on a variety of observations, the evolution of the universe can be reconstructed fairly accurately. We are currently living in a matter dominated universe $w=0$, and there is strong evidence for the presence of a positive cosmological constant $\Lambda >0$. Moreover, the spatial curvature of the universe appears to be zero $k=0$. There are some highly restrictive conditions in $k \neq 0$ models. 

Since the universe is currently expanding, it must have been smaller and denser in the past. From equations~(\ref{matter}) and~(\ref{rad}) we see that radiation decays faster than matter in an expanding universe. Therefore, at some point in the past, the universe was dominated by radiation. Going back in time further, the universe was very dense and therefore hot and relatively small. The `beginning' of the universe is often referred to as the big bang, giving the image of a vast explosion from which the evolution of the universe started.

It appears very likely that the universe also underwent a period of accelerated expansion at its very early stages, similar to the late time acceleration due to the cosmological term. The reasons for this are beyond the scope of this short introduction, however, we note that this epoch is called inflation. 

Very roughly speaking, the standard model of cosmology can be summarised by the succession of the following dominated eras
\begin{align}
  \text{inflation} \longrightarrow \text{radiation} 
  \longrightarrow \text{matter}
  \longrightarrow \text{cosmological term}
  \label{good}
\end{align}
and a good cosmological model should be able to reproduce (parts of) this pattern.

\subsection{A first taste of dynamical systems}
\label{sec:firsttaste}

In order to get a first taste of the usefulness of dynamical systems techniques in cosmology~\cite{dsincosmology,Coley:2003mj}, let us consider a universe which is spatially flat $k=0$, and its matter content is radiation $\rho_{\rm r}$ with $w=1/3$, and a perfect fluid (dust) $\rho_{\rm m}$ with $w=0$. The following four equations completely determine the dynamics of the system
\begin{subequations}
  \label{ex:fall}
  \begin{align}
    3H^2 - \Lambda &= \kappa\, (\rho_{\rm m} + \rho_{\rm r})
    \label{ex:f1}\\
    -2\dot{H} - 3H^ 2 + \Lambda&= \kappa\, \frac{1}{3} \rho_{\rm r} 
    \label{ex:f2}\\
    \dot{\rho}_{\rm r} + 4H\rho_{\rm r} &= 0
    \label{ex:f3}\\
    \dot{\rho}_{\rm m} + 3H\rho_{\rm m} &= 0.
    \label{ex:f4}
  \end{align}
\end{subequations}
Using the dimensionless density parameters $\Omega_{\rm m}$, $\Omega_{\rm r}$ and $\Omega_{\Lambda}$, we find that equation~(\ref{ex:f1}) becomes the constraint
\begin{align}
  1 = \Omega_{\rm m} + \Omega_{\rm r} + \Omega_{\Lambda} 
  \label{ex:c1}
\end{align}
which means that we have two independent quantities, and choose to work with $\Omega_{\rm m}$ and $\Omega_{\rm r}$. Moreover, since we expect energy densities to be positive we also have the conditions $0 \leq \Omega_{\rm m} \leq 1$ and $0 \leq \Omega_{\rm r} \leq 1$. Therefore, also $\Omega_\Lambda \leq 1$ is needed to satisfy equation~(\ref{ex:c1}). 

\begin{figure}[!htb]
  \centering
  \includegraphics[width=0.6\textwidth]{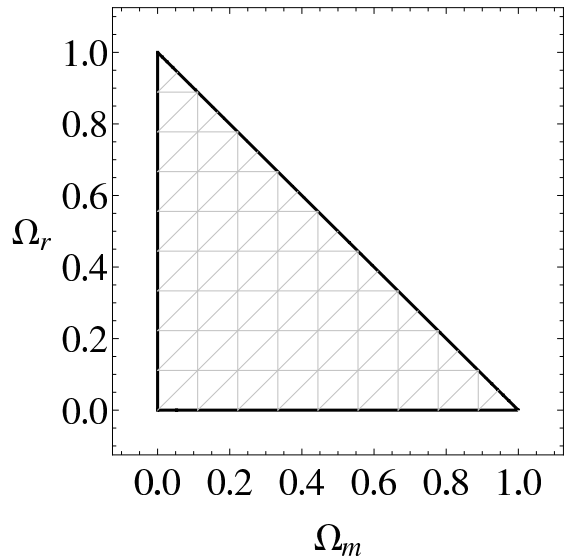}
  \caption{This figure shows the triangle defined by $\{(\Omega_{\rm m},\Omega_{\rm r})\, |\, 0 \leq \Omega_{\rm m} + \Omega_{\rm r} \leq 1 \cap 0 \leq \Omega_{\rm m} \leq 1 \cap 0 \leq \Omega_{\rm r} \leq 1\}$. Every solution to the field equations~(\ref{ex:fall}) corresponds to a trajectory inside this triangle, one calls this region the phase space of the system.}
\label{fig1}
\end{figure}

The solution to the system~(\ref{ex:fall}) at any given time $t$ will correspond to a point in the $(\Omega_{\rm m},\Omega_{\rm r})$ plane. The constraint equation~(\ref{ex:c1}) together with the aforementioned inequalities reduces the allowed $(\Omega_{\rm m},\Omega_{\rm r})$ plane to the triangle\footnote{To the best of our knowledge this idea goes back to Nicola Tamanini.} defined by $\Delta = \{(\Omega_{\rm m},\Omega_{\rm r})\, |\, 0 \leq \Omega_{\rm m} + \Omega_{\rm r} \leq 1 \cap 0 \leq \Omega_{\rm m} \leq 1 \cap 0 \leq \Omega_{\rm r} \leq 1\}$, see also Figure~\ref{fig1}.

Next, we wish to find the dynamical equations for the dimensionless variables $\Omega_{\rm m}$ and $\Omega_{\rm r}$. This requires a slightly lengthy but otherwise straightforward calculation of which we will show some details. We start with
\begin{align}
  \frac{d}{dt} \Omega_{\rm m} = 
  \frac{d}{dt} \left(\frac{\kappa\, \rho_{\rm m}}{3H^2}\right) = 
  \frac{\kappa}{3} \frac{\dot{\rho}_{\rm m} H^2 - \rho_{\rm m} 2 H \dot{H}}{H^4} =
  \frac{\kappa}{3 H} \left(\frac{\dot{\rho}_{\rm m}}{H} - 2 \rho_{\rm m} \frac{\dot{H}}{H^2}\right).
\end{align} 
From~(\ref{ex:f4}) we get an expression for $\dot{\rho}_{\rm m}/H$, while~(\ref{ex:f2}) can be solved for $\dot{H}/H^2$. This yields 
\begin{align}
  \frac{1}{H}\frac{d}{dt} \Omega_{\rm m} &= \frac{\kappa}{3 H^2} \left(
  -3\rho_{\rm m} + 3\rho_{\rm m} \bigl(1-\frac{\Lambda}{3H^2} + \frac{\kappa\, \rho_{\rm r}}{9H^2}\bigr)\right) \\
  &= -3 \Omega_{\rm m} + 3 \Omega_{\rm m}(1-\Omega_{\Lambda}+\Omega_{\rm r}/3).
\end{align} 
The last step is to eliminate $\Omega_{\Lambda}$ using~(\ref{ex:c1}) which gives the equation
\begin{align}
  \frac{1}{H}\frac{d}{dt} \Omega_{\rm m} &= 
  -3 \Omega_{\rm m} + 
  3 \Omega_{\rm m}(\Omega_{\rm m}+\Omega_{\rm r} + \Omega_{\rm r}/3) \\
  &= -3 \Omega_{\rm m} + 
  3 \Omega_{\rm m}(\Omega_{\rm m} + 4\Omega_{\rm r}/3) \\
  &= \Omega_{\rm m}(3\Omega_{\rm m} + 4\Omega_{\rm r} -3).
\end{align}
We now note that 
\begin{align}
  d \log(a) = \frac{\dot{a}}{a} dt = H dt
\end{align}
which means that by introducing the new independent variable $N = \log(a)$ and denoting differentiation with respect to $N$ by a prime, we finally arrive at
\begin{align}
  \Omega'_{\rm m} = \Omega_{\rm m}(3\Omega_{\rm m} + 4\Omega_{\rm r} -3).
  \label{ex:o1}
\end{align}
Following a similar calculation one can find the corresponding equation for $\Omega_{\rm r}$ which is given by
\begin{align}
  \Omega'_{\rm r} = \Omega_{\rm r}(3\Omega_{\rm m} + 4\Omega_{\rm r} -4).
  \label{ex:o2}
\end{align}
For any set of initial conditions $(\Omega_{\rm m}(N_i),\Omega_{\rm r}(N_i))$ with initial `time' $N_i$ in the triangle $\Delta$, the equations~(\ref{ex:o1}) and~(\ref{ex:o2}) will determine a trajectory which describes the dynamical behaviour of the cosmological model we are studying. It should be noted that equations~(\ref{ex:o1}) and~(\ref{ex:o2}) do not depend explicitly on the `time' parameter $N$, such a system is called an autonomous system of equations, or a dynamical system. Equations of this type can be studied using particular methods developed for such systems. In the next Section we will give a brief introduction to dynamical systems and the most common methods used to analyse them. 

\section{Some aspects of dynamical systems}

What is a dynamical system? It can be anything ranging from something as simple as a single pendulum to something as complex as the human brain and the entire universe itself. In general, a dynamical system can be thought of as any abstract system consisting of 
\begin{enumerate}
  \item a space (state space or phase space), and
  \item a mathematical rule describing the evolution of any point in that space. 
\end{enumerate}
The second point is crucial. Finding a mathematical rule which, for instance, describes the evolution of information at any neuron in the human brain is probably impossible. So, we need a mathematical rule as an input and finding one might be very difficult indeed.

The state of the system we are interested in is described by a set of quantities which are considered important about the system, and the state space is the set of all possible values of these quantities. In the case of the pendulum, the position of the mass and its momentum are natural quantities to specify the state of the system. For more complicated systems like the universe as a whole, the choice of good quantities is not at all obvious and it turns out to be useful to choose convenient variables. It is possible to analyse the same dynamical system with different sets of variables, either of which might be more suitable to a particular question. 

There are two main types of dynamical systems: The first are continuous dynamical systems whose evolution is defined by a set of ordinary differential equations (ODEs) and the other ones are called time-discrete dynamical systems which are defined by a map or difference equations. In the context of cosmology we are studying the Einstein field equations which for a homogeneous and isotropic space result in a system of ODEs. Thus we are only interested in continuous dynamical systems and will not discuss time-discrete dynamical systems in the remainder. 

Let us denote $\mathbf{x} = (x_1,x_2,\ldots,x_n) \in X$ to be an element of the state space $X\subseteq\mathbb{R}^n$. The standard form of a dynamical system is usually expressed as~\cite{wigginsbook}
\begin{align}
  \dot{\mathbf{x}} = \mathbf{f}(\mathbf{x})
  \label{dy1}
\end{align}
where the function $\mathbf{f}:X\rightarrow X$ and where the dot denotes differentiation with respect to some suitable time parameter. We view the function $\mathbf{f}$ as a vector field on $\mathbb{R}^n$ such that
\begin{align}
  \mathbf{f}(\mathbf{x}) = (f_1(\mathbf{x}),\cdots,f_n(\mathbf{x})).
\end{align}
The ODEs~(\ref{dy1}) define the vector fields of the system. At any point $x \in X$ and any particular time $t$, $\mathbf{f}(\mathbf{x})$ defines a vector field in $\mathbb{R}^n$. When discussing a particular solution to~(\ref{dy1}) this will often be denoted by $\psi(t)$ to simplify the notation. We restrict ourselves to systems which are are finite dimensional and continuous. In fact, we will require the function $f$ to be at least differentiable in $X$.

\begin{definition}[Critical point or fixed point]
  The autonomous equation $\dot{\mathbf{x}} = \mathbf{f}(\mathbf{x})$ is said to have a critical point or fixed point at $\mathbf{x} = \mathbf{x}_0$ if and only if $\mathbf{f}(\mathbf{x}_0) = 0$.
  \label{def1}
\end{definition}

There is an easy way to justify this definition. Let us consider a one dimensional mechanical system with force $F$. Newton's equation for such a system is
\begin{align}
  m \ddot{x} = F(x).
  \label{newton}
\end{align}
Let us introduce a second variable $p = m\dot{x}$ such that the single second order ODE~(\ref{newton}) becomes a system of two first order equations
\begin{align}
  \dot{x} &= p/m \\
  \dot{p} &= F(x).
  \label{newton2}
\end{align}
Therefore, according to Definition~\ref{def1}, the critical points of system~(\ref{newton2}) correspond to those points $x$ where the force vanishes $F(x)=0$. At these points, there is no force acting on the particle and the system could, in principle, remain in this (steady) state indefinitely.

This leads to the question of stability of a critical point or fixed point. The following two definitions will clarify what is meant by stable and asymptotically stable. In simple words a fixed point $x_0$ of the system~(\ref{dy1}) is called stable if all solutions $\mathbf{x}(t)$ starting near $\mathbf{x}_0$ stay close to it.

\begin{definition}[Stable fixed point]
  Let $\mathbf{x}_0$ be a fixed point of system~(\ref{dy1}). It is called stable if for every $\varepsilon > 0$ we can find a $\delta$ such that if $\psi(t)$ is any solution of~(\ref{dy1}) satisfying $\|\psi(t_0)-\mathbf{x}_0\| < \delta$, then the solution $\psi(t)$ exists for all $t \geq t_0$ and it will satisfy $\|\psi(t)-\mathbf{x}_0\| < \varepsilon$ for all $t \geq t_0$.
  \label{def2}
\end{definition}

The point is called asymptotically stable if it is stable and the solutions approach the critical point for all nearby initial conditions.

\begin{definition}[Asymptotically stable fixed point]
  Let $\mathbf{x}_0$ be a stable fixed point of system~(\ref{dy1}). It is called asymptotically stable if there exists a number $\delta$ such that if $\psi(t)$ is any solution of~(\ref{dy1}) satisfying $\|\psi(t_0)-\mathbf{x}_0\| < \delta$, then $\lim_{t \rightarrow \infty} \psi(t) = \mathbf{x}_0$.
  \label{def3}
\end{definition}

The main difference is simply that all trajectories near an asymptotically stable fixed point will eventually reach that point while trajectories near a stable point could for instance circle around that point. If the point is unstable then solutions will move away from it.

We will not encounter fixed points which are stable but not asymptotically stable when studying cosmological dynamical systems.

Having defined a concept of stability, we will now discuss methods which can be used to analyse the stability properties of critical points. 

\subsection{Linear stability theory}

The basic idea of linear stability theory can be explained neatly using the above one dimensional mechanical system $m \ddot{x} = F(x)$. Let us assume that there is a point $x_0$ where the force vanishes $F(x_0)=0$. Can we find the behaviour of the particle near this point? We set $x(t) = x_0 + \delta x(t)$ and assume $\delta x(t)$ to be small. Then $\ddot{x}(t) = \ddot{\delta x}(t)$ and $F(x) = F(x_0 + \delta x) \approx F(x_0) + F'(x_0) \delta x + \ldots = F'(x_0) \delta x + \ldots$ (recall $F(x_0)=0$) so that Newton's equations near the critical point becomes $m \ddot{\delta x} = F'(x_0) \delta x$ where $F'(x_0)$ is a constant. This is a linear second order constant coefficient ODE, its auxiliary equation is simply $\lambda^2 = F'(x_0)/m$. Therefore, the sign of $F'(x_0)$ determines the stability properties of the point $x_0$. If $F'(x_0) < 0$ the solution involves trigonometric functions and we would speak of a stable point, for $F'(x_0) > 0$ the solution would involve exponentials and we would refer to this point as unstable. 

Exactly the same ideas can be utilised when studying an arbitrary dynamical system. Let $\dot{\mathbf{x}} = \mathbf{f}(\mathbf{x})$ be a given dynamical system with fixed point at $\mathbf{x}_0$. We will now linearise the system around its critical point. Since $\mathbf{f}(\mathbf{x}) = (f_1(\mathbf{x}),\ldots,f_n(\mathbf{x}))$, we can Taylor expand each $f_i(x_1,x_2,\ldots,x_n)$ near $\mathbf{x}_0$
\begin{align}
  f_i(\mathbf{x}) = f_i(\mathbf{x}_0) + 
  \sum_{j=1}^{n} \frac{\partial f_i}{\partial x_j}(\mathbf{x}_0) y_j + 
  \frac{1}{2!} \sum_{j,k=1}^{n} \frac{\partial^2 f_i}{\partial x_j \partial x_k}(\mathbf{x}_0) y_j y_k + \ldots
\end{align}
where the vector $\mathbf{y}$ is defined by $\mathbf{y} = \mathbf{x} - \mathbf{x}_0$. Note that in what follows we are only interested in the first partial derivatives. Therefore, of particular importance is the object $\partial f_i/\partial x_j$ which if interpreted as a matrix is the Jacobian matrix of vector calculus of the vector valued function $\mathbf{f}$. We define
\begin{align}
  J = \frac{\partial f_i}{\partial x_j} =
  \begin{pmatrix} 
    \frac{\partial f_1}{\partial x_1} & 
    \ldots & 
    \frac{\partial f_1}{\partial x_n} \\
    \vdots & \ddots & \vdots \\
    \frac{\partial f_n}{\partial x_1} &
    \ldots & 
    \frac{\partial f_n}{\partial x_n}
  \end{pmatrix}
  \label{Jac}
\end{align}
It is the eigenvalues of the Jacobian matrix $J$, evaluated at the critical points $\mathbf{x}_0$, which contain the information about stability. In this context $J$ is sometimes referred to as the stability matrix of the system. As $J$ is an $n \times n$ matrix, it will have $n$, possibly complex, eigenvalues (counting repeated eigenvalues accordingly). Recalling the example of the one dimensional mechanical system at the beginning, it is clear that this approach might encounter problems if one or more of the eigenvalues are zero. This motivates the following definition~\cite{wigginsbook}.
\begin{definition}[Hyperbolic point]
  Let $\mathbf{x} = \mathbf{x}_0\in X \subset \mathbb{R}^n$ be a fixed point (critical point) of the system $\dot{\mathbf{x}} = \mathbf{f}(\mathbf{x})$. Then $x_0$ is said to be hyperbolic if none of the eigenvalues of the Jacobian matrix $J(\mathbf{x}_0)$ have zero real part. Otherwise the point is called non-hyperbolic.
\end{definition}
Linear stability theory fails for non-hyperbolic points and other methods have to be employed to study the stability properties.

Roughly speaking we are distinguishing three broad cases: If all eigenvalues have negative real parts, then we can regard the point as stable. If at least one eigenvalues has a positive real part, then the corresponding fixed point would not be stable and correspond to a saddle point which attracts trajectories in some directions but repels them along others. Lastly, all eigenvalues could have a positive real part, in which case all trajectories would be repelled. 

In more than 3 dimensions it becomes very difficult to classify all possible critical points based on their eigenvalues. However, in dimensions 2 and 3 this can be done. In the following we present all possible cases for two dimensional autonomous systems.

Let us consider the two dimensional autonomous system given by
\begin{subequations}
  \label{2dsys}
  \begin{align}
    \dot{x} &= f(x,y) \\
    \dot{y} &= g(x,y)
  \end{align}
\end{subequations}
where $f$ and $g$ are (smooth) functions of $x$ and $y$. We assume that there exits a hyperbolic critical point at $(x_0,y_0)$ so that $f(x_0,y_0)=0$ and $g(x_0,y_0)=0$. The Jacobian matrix of the system is given by
\begin{align}
  J = 
  \begin{pmatrix} 
    f_{,x} & f_{,y} \\
    g_{,x} & g_{,y}
  \end{pmatrix}
  \label{Jacobian:2d}
\end{align}
where the $f_{,x}$ means differentiation with respect to $x$. Its two eigenvalues $\lambda_{1,2}$ are given by
\begin{subequations}
\begin{align}
  \lambda_{1} &= 
  \frac{1}{2}(f_{,x}+g_{,y}) 
  + \frac{1}{2}\sqrt{(f_{,x}-g_{,y})^2 + 4 f_{,y}g_{,x}}\\
  \lambda_{2} &= 
  \frac{1}{2}(f_{,x}+g_{,y}) 
  - \frac{1}{2}\sqrt{(f_{,x}-g_{,y})^2 + 4 f_{,y}g_{,x}}
\end{align}
\end{subequations}
and be evaluated at any fixed point $(x_0,y_0)$.

Table~\ref{tab1} contains all possible cases in order to understand the stability or instability properties of the critical point $(x_0,y_0)$ based on the two eigenvalues $\lambda_1$ and $\lambda_2$.

\begin{table}[!htb]
  \centering
  \begin{tabular}{|p{0.25\textwidth}|p{0.65\textwidth}|}
    \hline
    Eigenvalues & Description \\
    \hline\hline
    $\lambda_1<0$, $\lambda_2<0$ & the fixed point is asymptotically stable and trajectories starting near that point will approach that point $\lim_{t\rightarrow \infty} (x(t),y(t)) = (x_0,y_0)$ \\
    \hline
    $\lambda_1>0$, $\lambda_2>0$ & the fixed point is unstable and trajectories will be repelled from the point $\lim_{t\rightarrow -\infty} (x(t),y(t)) = (x_0,y_0)$. We can speak of $(x_0,y_0)$ as the past time attractor \\
    \hline
    $\lambda_1<0$, $\lambda_2>0$ & the fixed point is a saddle point. Some trajectories will be repelled, others will be attracted \\
    \hline
    $\lambda_1 =0$, $\lambda_2>0$ & the point is unstable. The positive eigenvalues ensures that there is at least one unstable direction \\
    \hline
    $\lambda_1 =0$, $\lambda_2<0$ & linear stability theory fails to determine stability. The point is non-hyperbolic and other methods are needed to study the behaviour of trajectories near that point \\
    \hline
    $\lambda_1 = \alpha+i\beta$, $\lambda_2 = \alpha-i\beta$ & with $\alpha>0$ and $\beta\neq 0$ the fixed point is an unstable spiral \\
    \hline
    $\lambda_1 = \alpha+i\beta$, $\lambda_2 = \alpha-i\beta$ & with $\alpha<0$ and $\beta\neq 0$ the fixed point is a stable spiral \\
    \hline
    $\lambda_1 = i\beta$ , $\lambda_2 = -i\beta $ & solutions are oscillatory and the point is called a centre. Note that a critical point being a centre is not related to centre manifolds.\\
    \hline
  \end{tabular}
  \caption{Stability or instability properties of the critical point $(x_0,y_0)$ based on the two eigenvalues $\lambda_1$ and $\lambda_2$.}
  \label{tab1}
\end{table}

\subsection*{Example -- Cosmology with matter, radiation and cosmological term}

Recall the cosmological dynamical system~(\ref{ex:o1}) and~(\ref{ex:o2}) which will be our base model henceforth. The equations read
\begin{subequations}
  \label{ex:sys1}
  \begin{align}
    \Omega'_{\rm m} &= \Omega_{\rm m}(3\Omega_{\rm m} + 4\Omega_{\rm r} -3)\\
    \Omega'_{\rm r} &= \Omega_{\rm r}(3\Omega_{\rm m} + 4\Omega_{\rm r} -4)\\
    1 &= \Omega_{\rm m} + \Omega_{\rm r} + \Omega_{\Lambda}.
    \label{ex:sys1c}
  \end{align}
\end{subequations}
We can find the fixed points of this system by solving the simultaneous equations $\Omega'_{\rm m} = 0$ and $\Omega'_{\rm r} = 0$ for the pair $(\Omega_{\rm m},\Omega_{\rm r})$. We find three fixed points, namely $O=(0,0)$, $R=(0,1)$ and $M=(1,0)$. As we use the relative energy densities $\Omega_{i}$ as our dynamical variables, it is easy to interpret those fixed points. At $R$, the radiation dominates and normal matter is absent. Likewise, at $M$, the normal matter dominates while radiation is absent. The point $O$ contains neither radiation nor matter, and is therefore dominated by the cosmological term because of~(\ref{ex:sys1c}). 

\begin{figure}[!htb]
  \centering
  \includegraphics[width=0.8\textwidth]{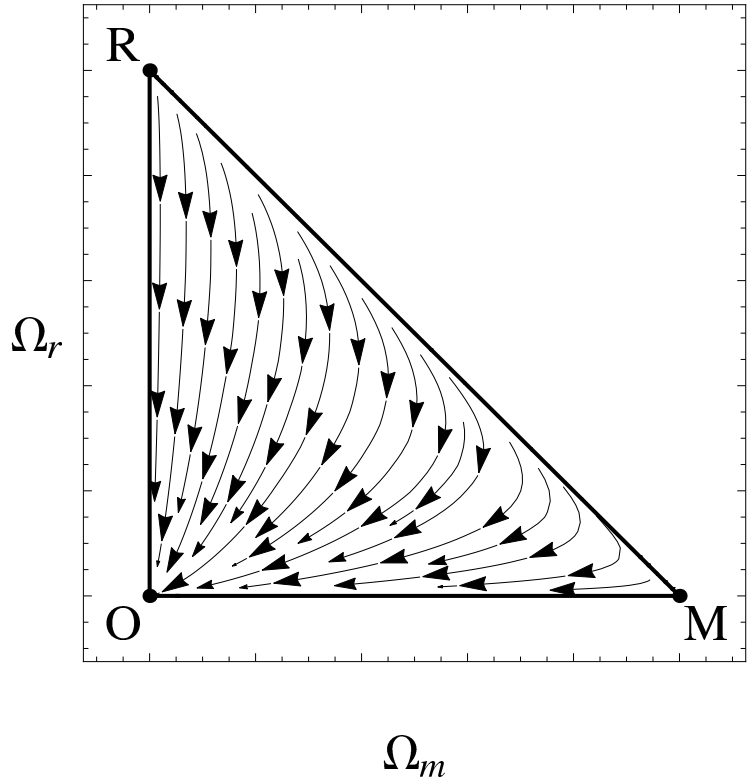}
  \caption{Phase space diagram of system~(\ref{ex:sys1}).}
\label{fig2}
\end{figure}

The Jacobian matrix of system~(\ref{ex:sys1}) is computed straightforwardly. Evaluated at the three fixed points, we find
\begin{align}
  J(O) = 
  \begin{pmatrix} 
    -3 & 0 \\
    0 & -4
  \end{pmatrix}, 
  \quad
  J(R) = 
  \begin{pmatrix} 
    1 & 0 \\
    3 & 4
  \end{pmatrix}, 
  \quad
  J(M) = 
  \begin{pmatrix} 
    3 & 4 \\
    0 & -1
  \end{pmatrix},
\end{align}
respectively. The corresponding eigenvalues of the stability matrix are given by
\begin{subequations}
  \begin{alignat}{3}
    &O: &\qquad \lambda_1 &= -3, &\quad \lambda_2 &= -4 \\
    &R: &\qquad \lambda_1 &= 1, &\quad \lambda_2 &= 4 \\
    &M: &\qquad \lambda_1 &= -1, &\quad \lambda_2 &= 3 
  \end{alignat}
\end{subequations}
which implies that that $O$ is the only attractor of the system. Therefore, all trajectories will eventually approach $O$. $R$ is unstable, however, since both eigenvalues are positive, we can think of $R$ as the only past time attractor. This means all trajectories will have `started' at $R$. Lastly, $M$ is a saddle point. This means that some trajectories are attracted towards $M$ but are eventually repelled to move towards $O$. The phase space diagram Fig.~\ref{fig2} clearly shows these features.

In the cosmological context this has the following interpretation. Consider a spatially flat universe filled with normal matter and radiation, and with a very small cosmological term\footnote{If the cosmological term happens to be `large' then matter will never dominate and one obtains an almost direct transition from radiation to a state where the cosmological term dominates.}. Such a universe will generically be dominated by radiation at early times, then it will undergo a period where matter dominates its energy contents. Eventually it will evolve to a state where the cosmological term dominates. This result is in line with our expectation of a good cosmological model, see~(\ref{good}).

\subsection{Lyapunov functions}

The following methods of studying the stability of a fixed point goes back to Lyapunov. It is completely different to linear stability and can be applied directly to the system in question. The main problem with this approach is that one has to be able to guess the Lyapunov function since there is no systematic way of doing so. Let us start by defining what a Lyapunov function is and its relation to stability of an autonomous system of equations.

\begin{definition}[Lyapunov function]
Let $\dot{\mathbf{x}} = \mathbf{f}(\mathbf{x})$ with $\mathbf{x} \in X \subset \mathbb{R}^n$ be a smooth autonomous system of equations with fixed point $\mathbf{x}_0$. Let $V : \mathbb{R}^n \rightarrow \mathbb{R}$ be a continuous function in a neighbourhood $U$ of $\mathbf{x}_0$, the $V$ is called a Lyapunov function for the point $\mathbf{x}_0$ if
\begin{enumerate}
\item $V$ is differentiable in $U \setminus \{\mathbf{x}_0\}$
\item $V(\mathbf{x}) > V(\mathbf{x}_0)$
\item $\dot{V} \leq 0 \quad \forall x\in U \setminus \{\mathbf{x}_0\}$.
\end{enumerate}
Note that the third requirement is the crucial one. It implies 
\begin{align}
  \frac{d}{dt} V(x_1,x_2,\ldots,x_n) & = \frac{\partial V}{\partial x_1} \dot{x}_1 + \ldots
  + \frac{\partial V}{\partial x_n} \dot{x}_n 
  \nonumber \\
  &= \frac{\partial V}{\partial x_1} f_1 + \ldots
  + \frac{\partial V}{\partial x_n} f_n \leq 0
\end{align}
which required repeated use of the chain rule and substitution of the autonomous system equations to eliminate the terms $\dot{x}_i$ for $i=1,\ldots,n$.
\end{definition}

One can conveniently write $dV/dt$ using vector calculus notation
\begin{align}
  \frac{d}{dt} V(x_1,x_2,\ldots,x_n) = \grad V \cdot \dot{\mathbf{x}} 
  = \grad V \cdot \mathbf{f}(\mathbf{x}).
\end{align}
Let us now state the main theorem which connects a Lyapunov function to the stability of a fixed point of a dynamical system.

\begin{theorem}[Lyapunov stability]\label{lyapunovtheorem}
  Let $\mathbf{x}_0$ be a critical point of the system $\dot{\mathbf{x}} = \mathbf{f}(\mathbf{x})$, and let $U$ be a domain containing $\mathbf{x}_0$. If there exists a Lyapunov function $V(\mathbf{x})$ for which $\dot{V} \leq 0$, then $\mathbf{x}_0$ is a stable fixed point. If there exists a Lyapunov function $V(\mathbf{x})$ for which $\dot{V} < 0$, then then $\mathbf{x}_0$ is a asymptotically stable fixed point. 

Furthermore, if $\|\mathbf{x}\| \rightarrow \infty$ and $V(\mathbf{x})\rightarrow\infty$ for all $\mathbf{x}$, then $\mathbf{x}_0$ is said to be globally stable or globally asymptotically stable, respectively.
\end{theorem}

One can also find some instability results, see e.g.~\cite{Brauer:1989}, which will also depend on our ability to find a suitable Lyapunov function. However, we will not use results along those lines since we are mainly concerned about the stability of certain fixed points in the context of cosmology.

Should we be able to find a Lyapunov function satisfying the criteria of the Lyapunov stability theorem, we could establish (asymptotic) stability without any reference to a solution of the ODEs. However, just because we failed in finding a Lyapunov function at a particular point does not necessarily imply that such a point is unstable. Since there is no systematic way of constructing a function, it is possible that we were simply not clever enough to find a Lyapunov function for the critical point concerned.

\subsubsection*{A first example}

This first example is taken from~\cite{wigginsbook}. Suppose that a system is described by the vector field
\begin{subequations}
  \begin{align}
    \dot{x} &= y\\
    \dot{y} &= -x + \epsilon x^2y
  \end{align}
\end{subequations}
which has one critical point at $(x,y)= (0,0)$. A candidate Lyapunov's function is given by
\begin{align}
  V(x,y) = \frac{x^2+y^2}{2},
\end{align}
satisfying $V(0,0)= 0$ and $V(x,y) > 0$ in the neighbourhood of the fixed point. This function leads to
\begin{align}
  \dot{V} = \grad V \cdot (\dot{x},\dot{y}) = \epsilon x^2 y^2
\end{align}
from which we conclude that the point is globally asymptotically stable if $\epsilon < 0$ since $x^2 y^2$ is positive definite and thus $\dot{V} < 0$ in the neighbourhood of the fixed point. It is important to emphasise, however, that $\epsilon > 0$ does not imply instability.

\subsubsection*{A second example}

Let us consider the system
\begin{subequations}
  \label{Lya:ex2}
  \begin{align}
    \dot{x} &= -x^3 + x y\\
    \dot{y} &= -y - 2x^2 - x^2 y
  \end{align}
\end{subequations}
which has one fixed point at $(x,y)= (0,0)$. Computing the eigenvalues of the Jacobian matrix at the fixed point yields $\lambda_1 = -1$ and $\lambda_2 = 0$. Therefore, we cannot decide, based on linear stability theory, whether the origin is stable or not. However, starting with the candidate Lyapunov function
\begin{align}
  V(x,y) = 2 x^2 + y^2
\end{align}
leads to
\begin{align}
  \dot{V} = -4x^4 - 2y^2 - 2x^2y^2.
\end{align}
Therefore the point is globally asymptotically stable since all terms in $\dot{V}$ are negative definite and thus $\dot{V} < 0$ in the neighbourhood of the fixed point. This example has been adapted from a similar one in~\cite{jackcarr}.

\begin{figure}[!htb]
  \centering
  \includegraphics[width=0.8\textwidth]{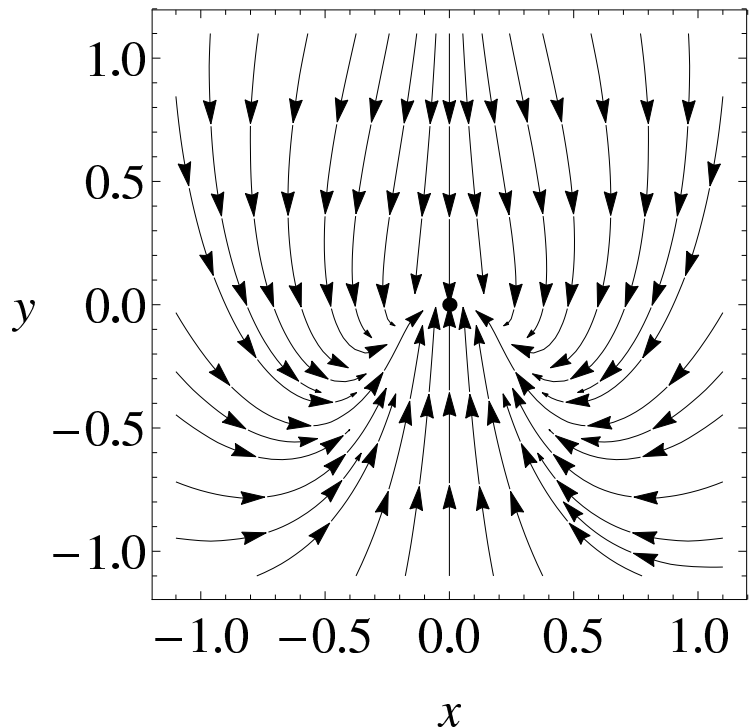}
  \caption{Phase space plot of the system~(\ref{Lya:ex2}).}
\label{fig3}
\end{figure}

Note that the phase space plot is in agreement with our conclusion of stability, see Fig.~\ref{fig3}.

\subsection{Centre manifold theory}

Centre manifold theory is a method that allows us to simplify dynamical systems by reducing their dimensionality near fixed points with vanishing eigenvalues of the Jacobian matrix. It is also central to other elegant concepts such as bifurcations and the method of normal forms~\cite{normalformsbook}. Here the essential basics of centre manifold theory are discussed following~\cite{wigginsbook} and~\cite{jackcarr}.

Let us, as above, consider the dynamical system
\begin{align}
  \dot{\mathbf{x}} = \mathbf{f}(\mathbf{x})
  \label{cdy1}
\end{align}
with $\mathbf{x} \in \mathbb{R}^n$ and let us assume that it has a fixed point $\mathbf{x}_0$. Near this point we can linearise the system using~(\ref{Jac}). Denoting $\mathbf{y}=\mathbf{x}-\mathbf{x}_0$, we can write~(\ref{cdy1})
\begin{align}
  \dot{\mathbf{y}} = J \mathbf{y}
  \label{cdy2}
\end{align}
where we emphasise that $J$ is a constant coefficient $n \times n$ matrix. As such it will have $n$ eigenvalues which motivates the following. The space $\mathbb{R}^n$ is the direct sum of three subspaces which are denoted by $\mathbb{E}^s$, $\mathbb{E}^u$ and $\mathbb{E}^c$, where the superscripts stand for stable, unstable and centre, respectively. The space $\mathbb{E}^s$ is spanned by the eigenvectors of $J$ which have negative real part, $\mathbb{E}^u$ is spanned by the eigenvectors of $J$ which have positive real part, and $\mathbb{E}^c$ is spanned by the eigenvectors of $J$ which have zero real part. Linear stability theory is sufficient to understand the dynamics of trajectories in $\mathbb{E}^s$ and $\mathbb{E}^u$. Centre manifold theory will determine the dynamics of trajectories in $\mathbb{E}^c$.

In the context of centre manifold theory it is useful to write our dynamical system~(\ref{cdy1}) in the form
\begin{subequations}
  \label{cmexvec}
  \begin{align}
    \dot{\mathbf{x}} &= A\mathbf{x} + f(\mathbf{x},\mathbf{y})
    \label{cenxdot}\\
    \dot{\mathbf{y}} &= B\mathbf{y} + g(\mathbf{x},\mathbf{y}),
    \label{cenydot}
  \end{align}
\end{subequations}
where $(x,y) \in\mathbb{R}^c\times\mathbb{R}^s$. Moreover, we assume
\begin{subequations}
  \label{cmnonlinearcondition}
  \begin{align}
    f(0,0) &= 0, \qquad \nabla f(0,0) = 0 \\
    g(0,0) &= 0, \qquad \nabla g(0,0) = 0.
  \end{align}
\end{subequations}
In the system~(\ref{cmexvec}), $A$ is a $c \times c$ matrix having eigenvalues with zero real parts, while $B$ is an $s \times s$ matrix whose eigenvalues have negative real parts. Our aim is to understand the centre manifold of this system in order to investigate its dynamics. We have suppressed some regularity assumptions on $f$ and $g$ for simplicity.

\begin{definition}[Centre Manifold]\label{cmdef}
  A geometrical space is a centre manifold for~(\ref{cmexvec}) if it can be locally represented as
  \begin{align}
    W^{c}(0) =\{(x,y)\in \mathbb{R}^c\times\mathbb{R}^s| y = h(x), |x|<\delta, h(0) = 0, \nabla h(0)= 0 \}
  \end{align}
  for $\delta$ sufficiently small.
\end{definition}
The conditions $h(0) = 0$ and $\nabla h(0) = 0$ from the definition imply that the space $W^c(0)$ is tangent to the eigenspace $E^c$ at the critical point $(x,y) = (0,0)$.

Centre manifold theory is based on three main theorems~\cite{wigginsbook}. The first one is about the existence of the centre manifold, the second one clarifies the issue of stability of solution while the last one is about constructing the actual centre manifold needed to investigate the stability. We will state those theorems but will not state the proofs, the interested reader is referred to~\cite{jackcarr}.

\begin{theorem}[Existence]\label{cmexist}
  There exists a centre manifold for~(\ref{cmexvec}). The dynamics of the system~(\ref{cmexvec}) restricted to the centre manifold is given by
  \begin{align}
    \dot{u} = Au + f(u,h(u))
    \label{cmexisteqn}
  \end{align}
  for $u\in\mathbb{R}^c$ sufficiently small.
\end{theorem}
\begin{theorem}[Stability]
  Suppose the zero solution of~(\ref{cmexisteqn}) is stable (asymptotically stable or unstable). Then the zero solution of~(\ref{cmexisteqn}) is also stable (asymptotically stable or unstable). Furthermore, if $(x(t),y(t))$ is also a solution of~(\ref{cmexisteqn}) with $(x(0),y(0))$ sufficiently small, there exists a solution $u(t)$ of~(\ref{cmexisteqn}) such that
  \begin{subequations}
    \begin{align}
      x(t) &= u(t) +\mathcal{O}(e^{-\gamma t})\\
      y(t) &= h(u(t)) + \mathcal{O}(e^{-\gamma t})
    \end{align}
  \end{subequations}
  as $t\rightarrow\infty$, where $\gamma>0$ is a constant.
\end{theorem}

We now know that the centre manifold exists, and we can establish the stability or instability of a solution. However, our ability to do so depends on the knowledge of the function $h(x)$ in Definition~\ref{cmdef}. We will now derive a differential equation for the function $h(x)$. 

Following Definition~\ref{cmdef}, we have that $y = h(x)$. Let us differentiate this with respect to time and apply the chain rule. This gives
\begin{align}
  \dot{y} = \nabla h(x) \cdot \dot{x}
  \label{cmydot}
\end{align}
Since $W^c(0)$ is based on the dynamics generated by the system~(\ref{cmexvec}), we can substitute for $\dot{x}$ the right-hand side of~(\ref{cenxdot}) and for $\dot{y}$ the right-hand side of~(\ref{cenydot}). This yields
\begin{align}
  Bh(x) + g(x,h(x)) = \nabla h(x) \cdot \left[Ax + f(x,h(x))\right]
\end{align}
where we also used that $y = h(x)$. The latter equation can be re-arranged into the quasilinear partial different equation
\begin{align}
  \mathcal{N}(h(x)) :=
  \nabla h(x)\left[Ax + f(x,h(x))\right] - Bh(x) - g(x,h(x)) = 0
  \label{cmn}
\end{align}
which must be satisfied by $h(x)$ for it to be the centre manifold. In general, we cannot find a solution to this equation. Even for relatively simple dynamical systems it is often impossible to find an exact solution of this equation. It is the third and last theorem which explain why not all is lost at this point.

\begin{theorem}[Approximation]\label{apptheorem}
  Let $\phi:\mathbb{R}^c\rightarrow\mathbb{R}^s$ be a mapping with $\phi(0) = \nabla \phi(0) = 0$ such that $\mathcal{N}(\phi(x)) = \mathcal{O}(|x|^q)$ as $x\rightarrow 0$ for some $q>1$. Then
  \begin{align}
    |h(x) - \phi(x)| = \mathcal{O}(|x|^q) \quad\text{as}\quad x\rightarrow 0.
  \end{align}
\end{theorem}

The main point of this theorem is that an approximate knowledge of the centre manifold returns the same information about stability as the exact solution of equation~(\ref{cmn}). It turns out that finding an approximation for the centre manifold is a fairly doable task in comparison to finding the exact solution. The centre manifold machinery is best explained with a concrete example.

\subsubsection*{Example -- a simple two-dimensional model}

The following two dimensional example is taken from Wiggins~\cite{wigginsbook}. We consider the system
\begin{subequations}
  \label{ex1}
  \begin{align}
    \dot{x} &= x^2y-x^5\\
    \dot{y} &= -y+x^2.
  \end{align}
\end{subequations}
The origin $(x,y) = (0,0)$ is a fixed point. The Jacobian matrix of the linearised system about the origin has eigenvalues of $0$ and $-1$. Since there is a zero eigenvalue, the point is non-hyperbolic and linear stability theory fails to determine the nature of stability of this point.

By Theorem~\ref{cmexist}, there exists a centre manifold for the system~(\ref{ex1}) and it can be represented locally as
\begin{align}
  W^c(0) = \{(x,y)\in\mathbb{R}^2|y=h(x),|x|<\delta, h(0) = Dh(0) = 0\}
\end{align}
for $\delta$ sufficiently small. Next, we need to compute $W^c(0)$. Here we can exploit Theorem~\ref{apptheorem} which says that it suffices to approximate the centre manifold to establish stability properties. Therefore, it is customary to assume an expansion for $h(x)$ of the form
\begin{align}
  h(x) = a x^2 + b x^3 + O(x^4)
  \label{hexpansion}
\end{align}
where $a$ and $b$ are constants to be determined. This expression is then substituted into~(\ref{cmn}) with the aim of determining those constants. 

\begin{figure}[!htb]
  \centering
  \includegraphics[width=0.8\textwidth]{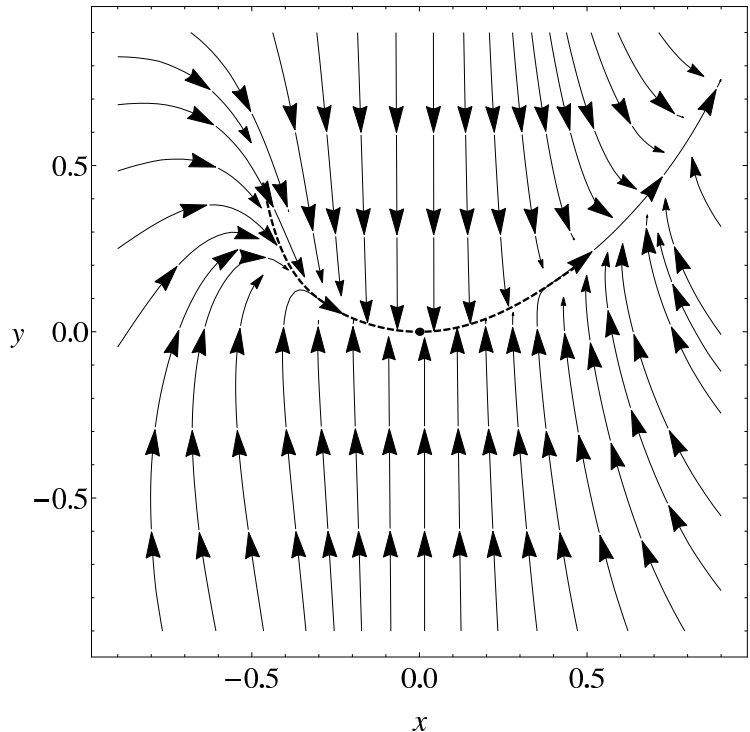}
  \caption{Phase space plot of the system~(\ref{ex1}). The centre manifold is indicated by a dashed line and was computed up to terms $x^{13}$. One sees very clearly how the centre manifold attracts the trajectories and how they are repelled from the origin (along the centre manifold) making this point unstable.}
\label{figcex}
\end{figure}

In this example, the equations~(\ref{ex1}) yield 
\begin{subequations}
  \begin{align}
    A &= 0 \quad B = -1 \\
    f(x,y) &= x^2y-x^5 \\
    g(x,y) &= x^2.
  \end{align}
\end{subequations} 
This, in addition to~(\ref{hexpansion}), is substituted into~(\ref{cmn}) and gives
\begin{align}
  \mathcal{N} &= (2ax + 3bx^2 + \cdots)(ax^4 + bx^5 - x^5 + \cdots)\nonumber\\
  &+ ax^2 + bx^3 - x^2 + \cdots = 0.
  \label{neqnexample}
\end{align}
The coefficients of each power of $x$ must be zero so that~(\ref{neqnexample}) holds. This provides us with a set on linear equations in the constants $a$ and $b$ which is solved by
\begin{align}
  a = 1 \quad b = 0,
\end{align}
where all terms of order $O(x^4)$ have been ignored. Therefore, the centre manifold is locally given by 
\begin{align}
  h(x) = x^2 + \mathcal{O}(x^4).
\end{align} 
Finally, following Theorem~\ref{cmexist}, the dynamics of the system restricted to the centre manifold is obtained to be
\begin{equation}
  \dot{x} = x^4 + \mathcal{O}(x^5).
  \label{cmexample1}
\end{equation}
We conclude that for $x$ sufficiently small, $x=0$ is unstable. Therefore, the critical point $(0,0)$ is unstable. In Fig.~\ref{figcex} we show the phase space for this system and also indicate the centre manifold.

\section{Cosmology using dynamical systems}

We discussed some aspects of cosmology in the context of dynamical systems, see~\cite{dsincosmology,Coley:2003mj} for more details, and also~\cite{Rendall:2001it}, or~\cite{Coley:2004jm} for anisotropic models. Based on the series of cosmological epochs $\text{inflation} \rightarrow \text{radiation} \rightarrow \text{matter} \rightarrow \text{cosmological term}$ of Section~\ref{sec:briefhist}, which could be called a `minimal' cosmological model, we will now make links with dynamical systems. A very neat paper studying cosmological models in the Lotka-Volterra framework is~\cite{Perez:2013zya}.

Let us now consider a generic `minimal' cosmological model described by an $n \times n$ system of autonomous equations. Should this model begin with an inflationary period, then this should correspond to an early time attractor in the dynamical system. All eigenvalues of the Jacobian matrix at this point should be positive in order to ensure that all trajectories evolve away from this point, this means $\lambda_i > 0$ for $i=1,\ldots,n$. 

In an ideal model we would also have two saddle points ($\lambda_j >0$, $\lambda_k <0$ with $j+k=n$) which correspond to a radiation dominated and matter dominated universe, respectively. These epochs being saddle points makes sure that some trajectories are attracted to these points, however, they will eventually be repelled. In this case the universe will evolve through both epochs. Let us note here that most models will only contain either matter or radiation, and thus we would be satisfied if there was only one saddle point.

Lastly, we require a late-time attractor ($\lambda_i < 0$ for $i=1,\ldots,n$) where the universe is undergoing an accelerated expansion which corresponds to the de Sitter solution. We say the universe is approaching de Sitter space asymptotically. This can be summarised as follows
\begin{align}
  \begin{array}{ccccc}
    \text{inflation} & 
    \longrightarrow & 
    \text{radiation/matter} &
    \longrightarrow & 
    \text{de Sitter}\\[1ex]
    \lambda_i > 0 & 
    \mbox{} & 
    \lambda_j >0, \lambda_k <0 &
    \mbox{} &
    \lambda_i < 0
  \end{array}
  \label{wishlist}
\end{align}
where, for simplicity, we neglected the possibility of some zero eigenvalues.

\subsection{Cosmology with matter and scalar field}

The cosmological constant $\Lambda$ has strong observational support~\cite{Perlmutter:1998,Riess:1998}, but also leads to a variety of problems which are called the cosmological constant problems, we refer the reader to~\cite{Weinberg:1988cp,Sahni:2002kh} and in particular~\cite{Martin:2012bt}. These problems can largely be avoided if the constant term $\Lambda$ is replaced by a dynamically evolving scalar field $\varphi$ with some given potential $V(\varphi)$. In this case one often speaks of dark energy. In many models the potential $V$ is assumed to be of exponential form, $V=V_0\exp(-\lambda\kappa\varphi)$.

Moreover, instead of writing the equation of state for the matter as $p=w\rho$, one often encounters a slightly different parametrisation which is given by
\begin{align}
  p_{\gamma} = w_{\gamma} \rho_{\gamma} = (\gamma-1)\rho_\gamma
\end{align}
where $\gamma = 1+w_{\gamma}$ is a constant and $0\leq\gamma\leq 2$. Its value is $4/3$ when there is radiation, and is $1$ for standard matter or dark matter in this context.

For this setup, the Einstein field equations are 
\begin{subequations}
  \label{eqn:sca1}
  \begin{align}
    H^2 &= \frac{\kappa^2}{3}
    \left(\rho_\gamma + \frac{1}{2}\dot{\varphi}^2 + V \right)
    \label{friedmann}\\
    \dot{H} &= -\frac{\kappa^2}{2}(\rho_\gamma+p_\gamma+\dot{\varphi}^2).
    \label{hduncoupled}
  \end{align}
\end{subequations}
We can interpret $\rho_\varphi = \dot{\varphi}^2/2+V$ as the energy density of the scalar field and $p_\varphi = \dot{\varphi}^2/2-V$ as its pressure. This also allows us to define an effective equation of state for the field. The conservation equations for the matter and the scalar field are given by
\begin{subequations}
  \label{eqn:sca2}
  \begin{align}
    \dot{\rho}_\gamma &= -3H (\rho_\gamma + p_\gamma)\\
    \ddot{\varphi} &= - 3H\dot{\varphi} - \frac{dV}{d\varphi} 
    = - 3H\dot{\varphi} + \lambda \kappa V
    \label{uncoupledKG}
  \end{align}
\end{subequations}
where we used the exponential form of the potential. We follow the approach outlined in Section~\ref{sec:firsttaste} and rewrite~(\ref{eqn:sca1}) and~(\ref{eqn:sca2}) using more suitable variables. As before, we start with dividing equation~(\ref{friedmann}) with $H^2$ which results in
\begin{align}
  1 = \frac{\kappa^2\rho_\gamma}{3H^2}+\frac{\kappa^2\dot{\varphi}^2}{6H^2}+\frac{\kappa^2V}{3H^2}.
  \label{friedmann2}
\end{align}
Every term on the right-hand side is positive since $V>0$ and $\rho_\gamma > 0$, and it turns out that the following the dimensionless variables~\cite{Copeland:1997et, Copeland:2006wr} are particularly useful
\begin{align}
  x^2 = \frac{\kappa^2\dot{\varphi}^2}{6H^2}, \quad
  y^2 = \frac{\kappa^2V}{3H^2}, \quad
  s^2 = \frac{\kappa^2\rho_\gamma}{3H^2}
  \label{compact}
\end{align}
which transform~(\ref{friedmann2}) into
\begin{align}
  1 = x^2 + y^2 + s^2.
  \label{friedmann3}
\end{align}
Therefore, we can choose $x,y$ as two independent variables. This leads to
\begin{align}
  1 \geq 1-x^2-y^2 = s^2 = \frac{\kappa^2\rho_\gamma}{3H^2} \geq 0
\end{align}
implying that $0\leq x^2+y^2\leq 1$ which means that the physical phase space of this model is contained within the unit circle.

We will introduce three more quantities which are useful in understanding the physical properties at the fixed points. The dimensionless density parameter~(\ref{density}) of the scalar field $\varphi$ can be expressed in terms of the new variables and is given by
\begin{align}
  \Omega_\varphi = \frac{\kappa^2\rho_\varphi}{3H^2} = x^2+y^2.
\end{align}
Moreover, we define the equation of state for the scalar field by
\begin{align}
  \gamma_\varphi = 1 + w_{\varphi} = 
  1 + \frac{p_{\varphi}}{\rho_{\varphi}} = \frac{2x^2}{x^2+y^2}.
\end{align}
Lastly, we define the effective equation of state of the total system by
\begin{align}
  w_{\rm eff} &= \frac{p_{\gamma} + p_{\varphi}}{\rho_{\gamma} + \rho_{\varphi}} =
  \frac{w_{\gamma}\rho_{\gamma} + \dot{\varphi}^2/2-V}{\rho_{\gamma} + \dot{\varphi}^2/2+V} 
  \nonumber \\
  &= w_{\gamma}(1-x^2-y^2) + x^2 - y^2.
\end{align}

Now, we are ready to derive a two dimensional dynamical system using the variables $x$ and $y$. As before, we will introduce a new `time' variable $N=\log(a)$ so that $dN = H dt$, and denote differentiation with respect to $N$ by a prime.

Let us begin by differentiating $x$ with respect to time $t$
\begin{align}
  \dot{x} = \frac{\kappa}{\sqrt{6}} \frac{\ddot{\varphi}H -\dot{\varphi}\dot{H}}{H^2} = 
  \frac{\kappa}{\sqrt{6}}\left(\frac{\ddot{\varphi}}{H}-\dot{\varphi}\frac{\dot{H}}{H^2}\right).
\end{align}
Substituting for $\ddot{\varphi}$ using~(\ref{uncoupledKG}) and for $\dot{H}$ using~(\ref{hduncoupled}) we arrive at
\begin{align}
  \dot{x} = \frac{\kappa}{\sqrt{6}}\left(- 3\dot{\varphi} + \lambda\kappa \frac{V}{H} + \dot{\varphi}\frac{\kappa^2}{2H^2}(\gamma\rho_\gamma+\dot{\varphi}^2)\right).
\end{align}
Next, using the variables~(\ref{compact}) and the condition~(\ref{friedmann3}) we get
\begin{align}
  \dot{x} = H\left[-3x +\sqrt{\frac{3}{2}}\lambda y^2 + \frac{3}{2}x\left((1-x^2-y^2)\gamma + 2x^2\right)\right]
  \label{finalxdot}
\end{align}
One can now introduce the new `time parameter' $N$. Following similar steps, the equation for $y'$ can be derived. The final system is
\begin{subequations}
  \label{xyp}
  \begin{align}
    x' &= -3x + \sqrt{\frac{3}{2}}\lambda y^2 + \frac{3}{2}x\left(2x^2 + \gamma(1-x^2-y^2)\right)\\
    y' &= -\lambda\sqrt{\frac{3}{2}}xy + \frac{3}{2}y\left(2x^2 + \gamma(1-x^2-y^2)\right).
  \end{align}
\end{subequations}
The complete dynamics of this cosmological model is describe by the two equations~(\ref{xyp}).

We noted that the phase space of this system is contained in the unit circle. Inspection of the dynamical equations shows that system~(\ref{xyp}) is invariant under the transformation $y \mapsto -y$ and symmetric under time reversal $t \mapsto -t$. This implies that we can restrict our analysis on the upper half-disk with $y > 0$. The lower half-disc of the phase space corresponds to the contracting universe because $H<0$ in this region.

The properties of the dynamical system~(\ref{xyp}) depend on the values of the constants $\lambda$ and $\gamma$. Amongst others, they will in particular affect the existence and stability of the fixed points of the system, see~\cite{Copeland:1997et}. This can be related to the theory of bifurcations, something that has not been explored in cosmological dynamical systems. The following Table~\ref{crit1} contains all critical points of the system~(\ref{xyp}).

\begin{table}[!htb]
  \centering
  \begin{tabular}{|c|c|c|c|}
    \hline
    \mbox{} & $x$ & $y$ & existence \\
    \hline 
    \hline 
    O & 0 & 0 & $\forall\lambda$ and $\gamma$ \\
    \hline
    $\mathrm{A}_{+}$ & 1 & 0 & $\forall\lambda$ and $\gamma$ \\
    \hline
    $\mathrm{A}_{-}$ & -1 & 0 & $\forall\lambda$ and $\gamma$ \\
    \hline
    B & $\lambda/\sqrt{6}$ & $[1-\lambda^2/6]^{1/2}$ & $\lambda^2 < 6$ \\
    \hline
    C & $\sqrt{3/2} \gamma/\lambda$ & $[3(2-\gamma)\gamma/2\lambda^2]^{1/2}$ 
    & $\lambda^2 > 3\gamma$ \\
    \hline
  \end{tabular}
  \caption{Critical point of the system~(\ref{xyp}).}
  \label{crit1}
\end{table}

Having found all the possible fixed points, we can now compute the eigenvalues and determine their stability which is summarised in Table~\ref{crit2}, see~\cite{Copeland:1997et}.

\begin{table}[!htb]
  \centering
  \begin{tabular}{|c|p{0.6\textwidth}|c|c|}
    \hline
    \mbox{} & Stability & $\Omega_\varphi$ & $\gamma_\varphi$ \\ 
    \hline 
    \hline 
    O & saddle point for $0 < \gamma < 2$ & 0 & Undefined \\
    \hline
    $\mathrm{A}_{+}$ & unstable node for $\lambda < \sqrt{6}$ and saddle point for $\lambda > \sqrt{6}$& 1 & 2 \\ 
    \hline
    $\mathrm{A}_{-}$ & unstable node for $\lambda >-\sqrt{6}$ and addle point for $\lambda < -\sqrt{6}$ & 1 & 2 \\ 
    \hline
    B & stable node for $\lambda^2 < 3\gamma$ and saddle point for $3\gamma < \lambda^2 < 6$ & 1 & $\lambda^2/3$ \\ 
    \hline
    C & stable node for $3\gamma < \lambda^2 < 24 \gamma^2/(9\gamma -2)$ and stable spiral for $\lambda^2 > 24 \gamma^2/(9\gamma -2)$ & $3\gamma/\lambda^2$ & $\gamma$ \\ 
    \hline
\end{tabular}
\caption{Summary of the properties of the critical points.}
\label{crit2}
\end{table}

\begin{figure}[!htb]
  \centering
  \includegraphics[width=0.9\textwidth]{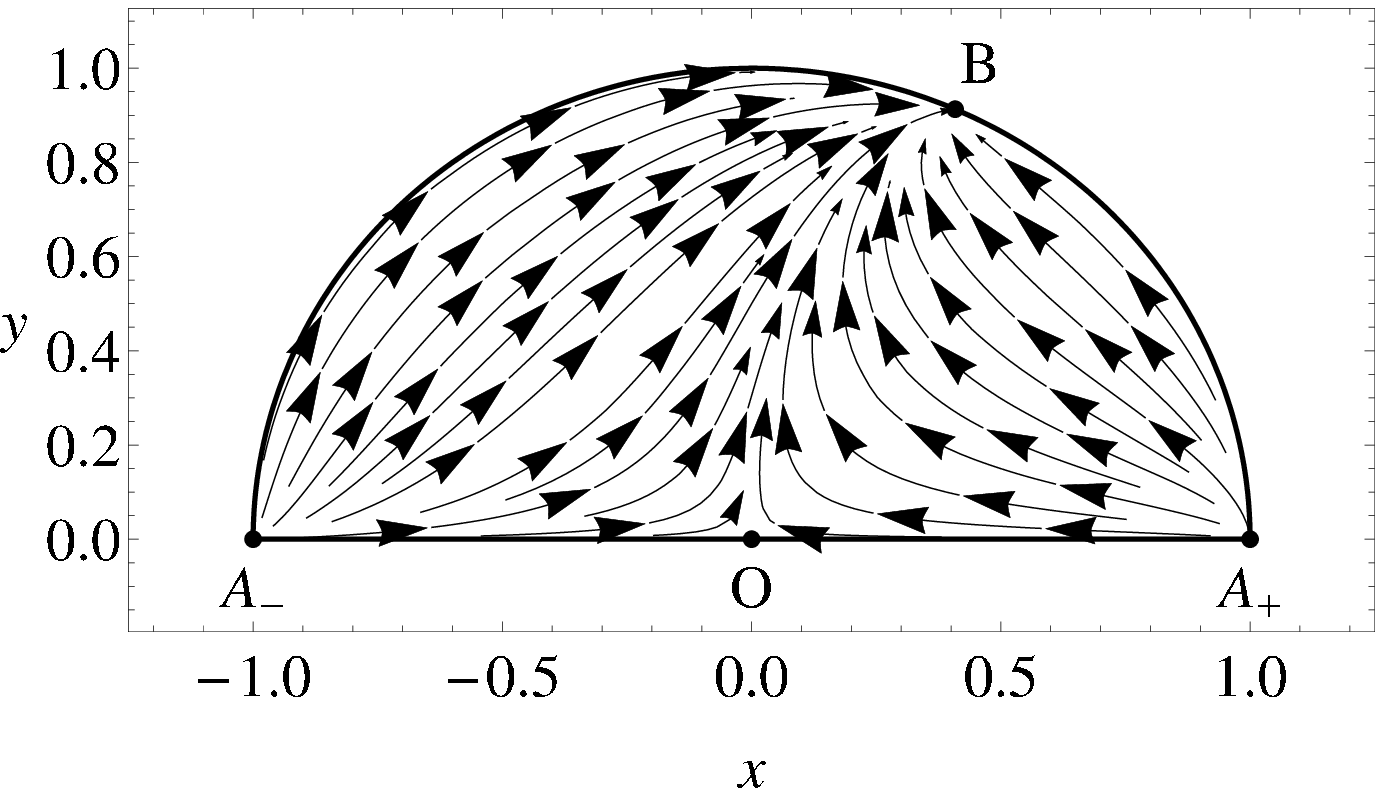}
  \caption{Phase space plot scalar field cosmology with exponential potential and matter. Parameter values are $\gamma=1$ and $\lambda=1$.}
\label{fig_d1}
\end{figure}

\begin{figure}[!htb]
  \centering
  \includegraphics[width=0.9\textwidth]{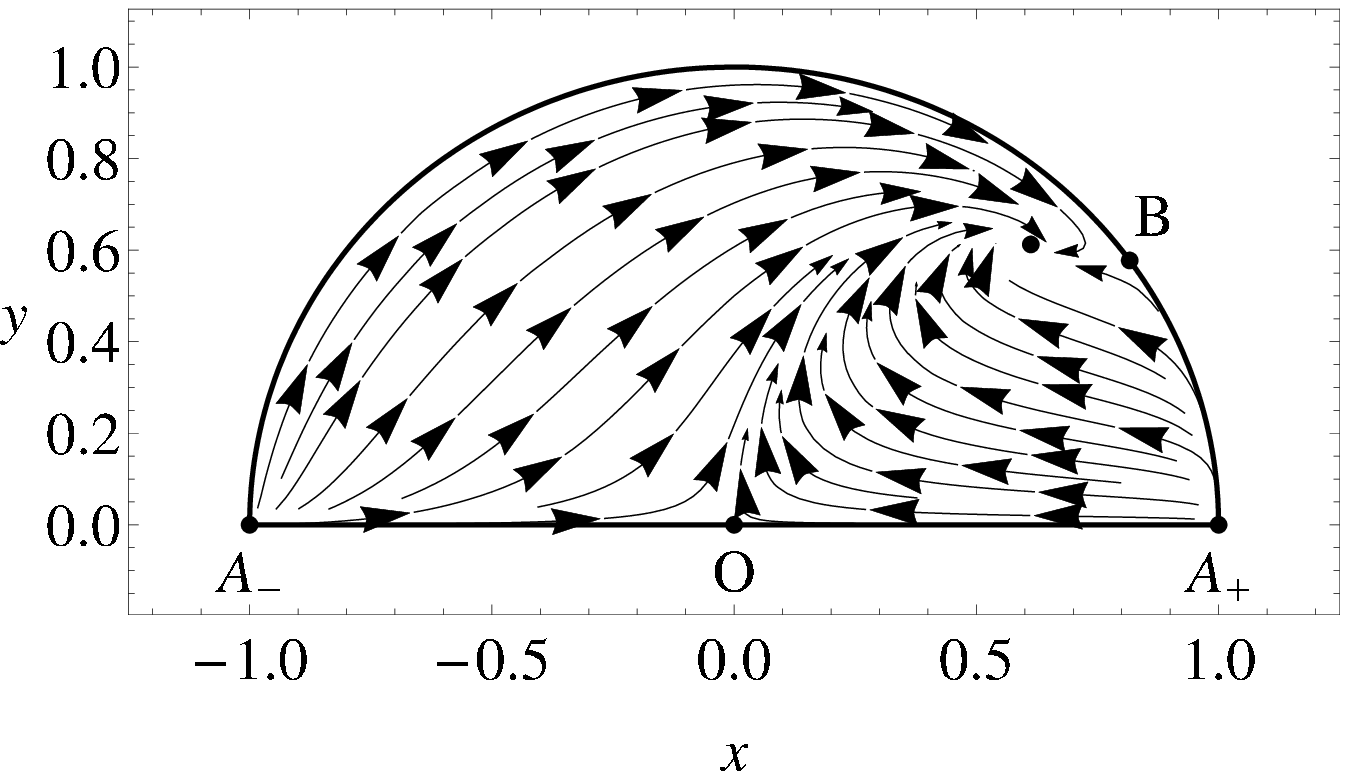}
  \caption{Phase space plot scalar field cosmology with exponential potential and matter. Parameter values are $\gamma=1$ and $\lambda=2$.}
\label{fig_d2}
\end{figure}

\begin{figure}[!htb]
  \centering
  \includegraphics[width=0.9\textwidth]{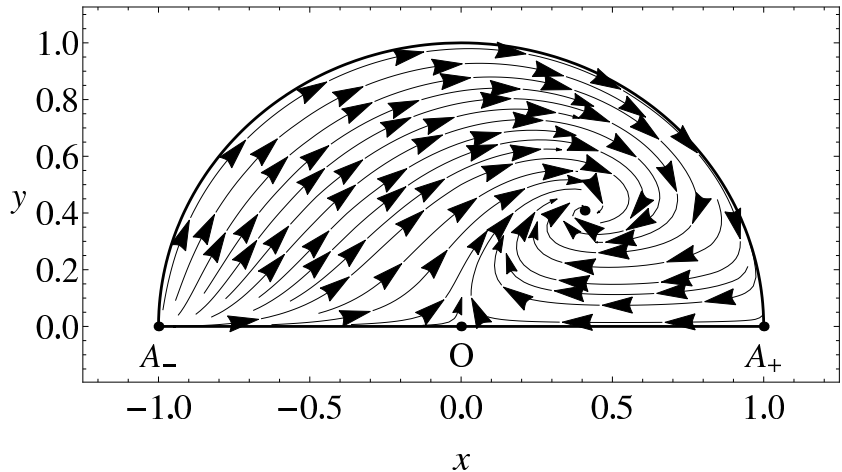}
  \caption{Phase space plot scalar field cosmology with exponential potential and matter. Parameter values are $\gamma=1$ and $\lambda=3$.}
\label{fig_d3}
\end{figure}

The three figures Fig.~\ref{fig_d1}--\ref{fig_d3} show the phase spaces of this model for various parameter choices.

It should be noted that the inequality signs in Table~\ref{crit2} exclude certain values from the analysis. For instance, when we choose $\lambda^2 = 3\gamma$, the two points B and C have the same coordinates (the system has one critical point less), namely $x_0 = \sqrt{\gamma/2}$ and $y_0 = \sqrt{1-\gamma/2}$ so that $x_0^2+y_0^2=1$ and its eigenvalues are $0,3/2(\gamma-2)$. Linear stability theory cannot determine the stability of this point. One could, in principle, apply centre manifold theory. However, this is problematic as the physical phase space is bounded by the unit circle and centre manifold theory will take into account the entire phase space. One could construct the centre manifold and only consider it inside the circle but this also has problems. For concreteness we set $\gamma=1$ in the following, which means $\lambda = \sqrt{3}$ and $x_0 = y_0 = \sqrt{1/2}$.

\begin{figure}[!htb]
  \centering
  \includegraphics[width=0.9\textwidth]{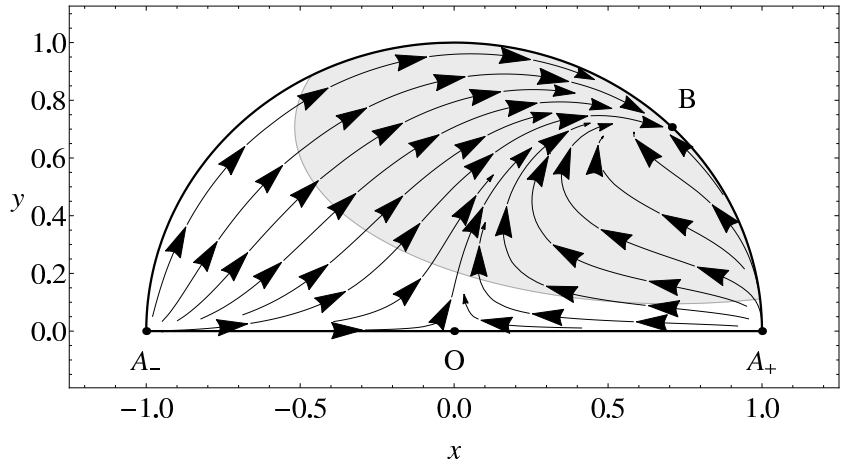}
  \caption{Phase space plot scalar field cosmology with exponential potential and matter. Parameter values are $\gamma=1$ and $\lambda=\sqrt{3}$. The shaded area shows part of the region of asymptotic stability of the fixed point. In this region $\dot{V} < 0$ and $V < 3/2$.}
\label{fig_d4}
\end{figure}

The easiest way forward is to use Lyapunov's method near this point. We start with the candidate Lyapunov function of the form
\begin{align}
  V = \left(x-\frac{1}{\sqrt{2}}\right)^2 + 4 \left(y-\frac{1}{\sqrt{2}}\right)^2
  \label{Vdyn}
\end{align}
and one verify that this function satisfies $\dot{V} < 0$ near the critical point. Since the function is positive definite near that point by construction, we can apply Theorem~\ref{lyapunovtheorem}. Following for instance~\cite{Brauer:1989}, we can estimate the region of asymptotic stability. Defining $S_\delta := \{(x,y)| V \leq \delta \}$ for $\delta \geq 0$, and denoting by $C_\delta$ the component of $S_\delta$ containing the critical point, we have the following statement~\cite{Brauer:1989}. Let $\Omega$ be the set where $\dot{V} < 0$, then the interior of $C_\delta$ contained in $\Omega$ lies in the region of asymptotic stability. As mentioned earlier, this approach relies on our ability to find a suitable Lyapunov function. Different choices can result in different parts of the region of asymptotic stability being covered and there is no guarantee that the entire region can be identified by this method alone. In Fig.~\ref{fig_d4}, we show the region of asymptotic stability based on the Lyapunov function~(\ref{Vdyn}) for model~(\ref{xyp}). A better Lyapunov function would of course improve this picture and increase the region.

A detailed and comprehensive phase-space analysis, based on linear stability theory alone, of this model can be found in~\cite{Copeland:1997et}. Other methods were explored in~\cite{Bohmer:2010re}. A complete discussion of all its properties in the context of cosmology is also given. This model has many interesting features as well as some problems which motivates various extensions, many of which have been considered in the literature. In fact, the literature of dynamical systems applications in early-time and late-time cosmology is so vast, that it could fill several books with ease! 

We should point out that this model falls short our wish list~(\ref{wishlist}). The early time fixed points $\mathrm{A}_{\pm}$ are dominated by the scalar field, however, the effective equation of state is $w_{\rm eff} = 1$ which is unphysical. It is point C which makes this model so interesting because this fixed point is stable and contains both, a non-vanishing scalar field and matter. One speaks of scaling solutions as the scalar field energy density is proportional to that of the fluid.

\subsection{Cosmology with matter and scalar field and interactions}

The models considered so far were all two dimensional. This relied on the fact that we were able to `eliminate' the Hubble parameter $H$ from the equations due to a smart choice of variables and a clever choice of `time'. However, there are many known models where this approach does not work and one has to introduce new variables. In the following we will discuss one such type of models and a possible choice of a new variable. 

The cosmological Einstein field equations~(\ref{eqn:sca1})--(\ref{eqn:sca2}) are compatible with the introduction of an additional interaction term $Q$, say. This interaction would allow for an energy transfer from the scalar $\varphi$ to the matter $\rho_\gamma$ and vice versa. The introduction of such a term leaves Eqs.~(\ref{eqn:sca1}) unchanged, but~(\ref{eqn:sca2}) becomes
\begin{subequations}
  \label{eqn:sca3}
  \begin{align}
    \dot{\rho}_\gamma &= -3H (\rho_\gamma + p_\gamma) - Q\\
    \ddot{\varphi} &= - 3H\dot{\varphi} - \frac{dV}{d\varphi} + \frac{Q}{\dot{\varphi}}
    \label{coupledKG}
  \end{align}
\end{subequations}
where we note that the term $Q\dot{\varphi}$ is natural when one computes the conservation equation $\dot{\rho}_\varphi = -3H (\rho_\varphi+p_\varphi)$. Various choices for the coupling function $Q$ were considered in the literature, for instance $Q=\alpha H\rho_\gamma$ or $Q=(2/3)\kappa\beta\rho_\gamma\dot{\varphi}$ with $\alpha$ and $\beta$ being dimensionless constants whose sign determines the direction of energy transfer from one component to the other~\cite{Amendola:1999qq, Holden:1999hm, Billyard:2000bh}. Those two choices can be motivated physically, however, one of the main motivation is the fact that the dynamical system with these coupling remains two dimensional as the Hubble parameter can be eliminated from the equations. However, both choices appear rather arbitrary and one would prefer a choice where the coupling is simply proportional to an energy density, for instance $Q = \Gamma \rho_{\gamma}$ with $\Gamma$ assumed to be small, see~\cite{Boehmer:2008av}, or for a further generalisation~\cite{Boehmer:2009tk}. In this case the phase space cannot be represented in the plane and one has to work in a three dimensional space.

As before, we start with the variables~(\ref{compact}) but need a third variable in order to be able to write the cosmological field equations as an autonomous system of differential equations. A possible third variable $z$ can be chosen to be
\begin{align}
  z = \frac{H_0}{H+H_0}
\end{align}
where $H_0$ is the Hubble parameter at an arbitrary fixed time. It is convenient to chose this time to be `today'. This variable $z$ ensures that the physical phase-space is compact. The Hubble parameter $H \rightarrow 0$ in the early time universe and $H \rightarrow \infty$ for the late time universe. Therefore 
\begin{align} 
  z = \begin{cases} 
    0 &\mbox{if } H=0 \\ 
    1/2 &\mbox{if } H=H_0 \\
    1 &\mbox{if } H \rightarrow \infty
  \end{cases}
\end{align}
and $z$ is bounded by $0 \leq z \leq 1$. Since the phase-space of system~(\ref{xyp}) is half a unit circle, we have that with coupling term $Q = \Gamma \rho_{\gamma}$ the phase-space now corresponds to a half-cylinder of unit height and unit radius.

\begin{figure}[!htb]
  \centering
  \includegraphics[width=0.6\textwidth]{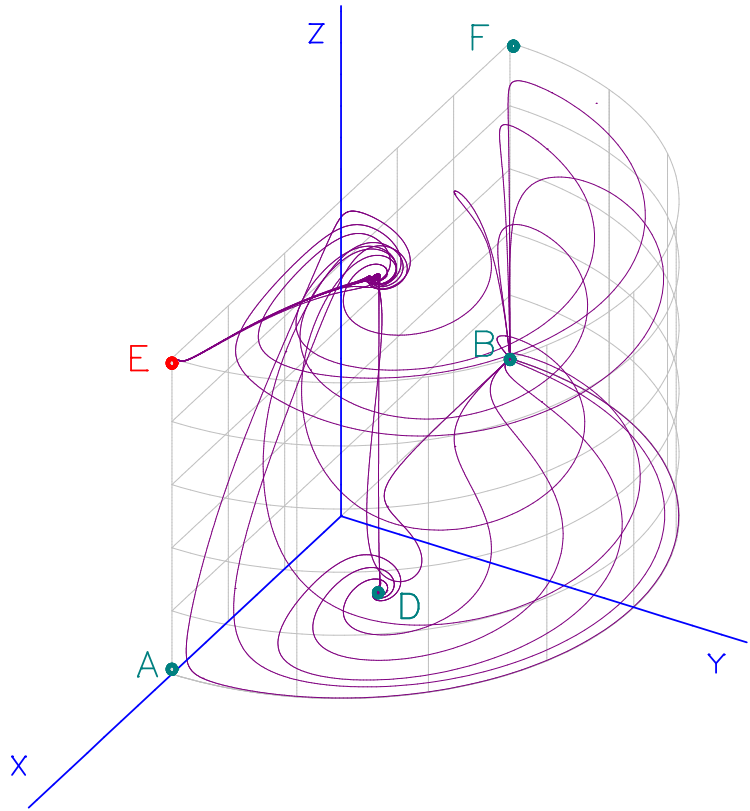}\\[3em]\mbox{}
  \caption{Phase space plot of system~(\ref{withcoup}) with $\lambda=4$ and $\zeta = 10^{-6}$.}
\label{fig_ql4}
\end{figure}

The resulting dynamical system is given by
\begin{subequations}
  \label{withcoup}
  \begin{align}
    \label{x'}
    x' &= -3x+\lambda \frac{\sqrt{6}}{2}\,y^2+\frac{3}{2}x(1+x^2-y^2)
    -\zeta\,\frac{(1-x^2-y^2)z}{2x(z-1)} \\
    \label{y'}
    y' &= -\lambda \frac{\sqrt{6}}{2}\,xy+\frac{3}{2}y(1+x^2-y^2) \\
    \label{z'}
    z' &= \frac{3}{2}z(1-z)(1+x^2-y^2)
    \end{align}
\end{subequations}
where $\zeta = \Gamma/H_0$. A detailed phase-space analysis of this model can be found in~\cite{Boehmer:2008av}. However, this model also has some additional interesting features outside the standard linear stability theory. For instance, there is a point vertically above point D in Fig.~\ref{fig_ql4} which attract trajectories. The system~(\ref{withcoup}), however, does not have a critical point there. 

By inspecting equations~(\ref{withcoup}) at $x_0=y_0=\sqrt{6}/(2\lambda)$ we note that $y'(x_0,y_0) = 0$ and that
\begin{subequations}
  \label{withcoup2}
  \begin{align}
    x'(x_0,y_0) &= -\zeta\,\frac{(\lambda-3/\lambda)z}{\sqrt{6}(z-1)} \\
    z'(x_0,y_0) &= \frac{3}{2}z(1-z)
  \end{align}
\end{subequations}
for some small coupling term $\zeta \ll 1$. Therefore, as the trajectories approach the $z=1$ plane, we have that also $z'(x_0,y_0) \rightarrow 0$. However, the behaviour of $x'(x_0,y_0)$ is more involved as
\begin{align}
    x'(x_0,y_0) \propto \frac{\zeta}{1-z}.
\end{align}
On the other hand, $\zeta \ll 1$ but $(1-z) \rightarrow 0$ as $z \rightarrow 1$. Therefore, the fraction $\zeta/(1-z)$ will initially be small. However, eventually the $(1-z)$ term will dominate and $\zeta/(1-z)$ will become large, explaining the repeller behaviour of this point in Fig.~\ref{fig_ql4}.  

\section{Final remarks}

It is hoped that this chapter succeeded in giving the reader a useful introduction into the exciting field of dynamical systems in cosmology. We should remark that the majority of papers dealing with the subject are confined to linear stability theory and focus more on the interpretation of results in the context of cosmology. However, there are many models where a more in depth analysis is needed to gain a complete understanding of the physics involved. Moreover, there is no need to select models primarily because of their simpler mathematical structure since we have all the tools at hand to study the more difficult ones too. We hope the reader feels encouraged to study all aspects of a cosmological dynamical system and use a variety of techniques developed by mathematicians, beyond linear stability theory. As Einstein wrote `Everything should be made as simple as possible, but not simpler.'

\subsection*{Acknowledgements}
We would like to thank Nicola Tamanini and Matthew Wright for valuable comments on these notes.


\end{document}